\documentclass[%
 aps%
 ,prl%
 ,amsmath,amssymb
 ,
 ,reprint%
 ,superscriptaddress
,
,floatfix
]{revtex4-1}

\usepackage{graphicx}
\usepackage{siunitx}
\usepackage{setspace}
\usepackage{enumitem}
\makeatletter
\newcommand{\manualsublabel}[3]{(#2)\def\@currentlabel{\ref{#3}(#2)}\label{#1}}
\makeatother

\begin{document}
\title{Momentum-dependent power law measured in an interacting quantum wire beyond the Luttinger limit}



\author{Y. Jin}
\affiliation{Department of Physics, Cavendish Laboratory, University of Cambridge, Cambridge, CB3 0HE, UK} 

\author{O. Tsyplyatyev}%
\email[E-mail: ]{o.tsyplyatyev@gmail.com.}
\affiliation{Institut f\"ur Theoretische Physik, Universit\"at Frankfurt,
Max-von-Laue Stra{\ss}e 1, 60438 Frankfurt, Germany}

\author{M. Moreno}%
\affiliation{Department of Physics, Cavendish Laboratory, University of Cambridge, Cambridge, CB3 0HE, UK}

\author{A. Anthore}
\affiliation{Universit\'{e} Paris Diderot, Sorbonne Paris Cit\'{e}, 75013 Paris, France}

\author{W.~K. Tan}%
\affiliation{Department of Physics, Cavendish Laboratory, University of Cambridge, Cambridge, CB3 0HE, UK}

\author{J.~P. Griffiths}%
\affiliation{Department of Physics, Cavendish Laboratory, University of Cambridge, Cambridge, CB3 0HE, UK}

\author{I. Farrer}
\altaffiliation[Present address: ]{Department of Electronic \& Electrical Engineering, University of Sheffield, 3 Solly Street, Sheffield, S1 4DE, UK.}
\affiliation{Department of Physics, Cavendish Laboratory, University of Cambridge, Cambridge, CB3 0HE, UK}

\author{D.~A. Ritchie}
\affiliation{Department of Physics, Cavendish Laboratory, University of Cambridge, Cambridge, CB3 0HE, UK}

\author{L.~I. Glazman}
\affiliation{Departments of Physics and Applied Physics, Yale University, New Haven, CT 06520, USA}

\author{A.~J. Schofield}%
\affiliation{School of Physics and Astronomy, University of Birmingham, Edgbaston, Birmingham, B15 2TT, UK}

\author{C.~J.~B. Ford}
\email[E-mail: ]{cjbf@cam.ac.uk.}
\affiliation{Department of Physics, Cavendish Laboratory, University of Cambridge, Cambridge, CB3 0HE, UK}

\begin{abstract}

Power laws in physics have until now always been associated with a scale invariance originating from the absence of a length scale. Recently, an emergent invariance even in the presence of a length scale has been predicted by the newly-developed nonlinear-Luttinger-liquid theory for a one-dimensional (1D) quantum fluid at finite energy and momentum, at which the particle's wavelength provides the length scale. 
We present the first experimental example of this new type of power law in the spectral function of interacting electrons in a quantum wire using a transport-spectroscopy technique. The observed momentum dependence of the power law in the high-energy region matches the theoretical predictions, supporting not only the 1D theory of interacting particles beyond the linear regime but also the existence of a new type of universality that emerges at finite energy and momentum.
\end{abstract}

\maketitle

Power laws play an important role in physics and they are generally associated with a scale invariance originating from the absence of a length scale. The most notable example is in continuous phase transitions where the diverging correlation length means that microscopic details become irrelevant and universality classes characterise the exponents\cite{Widom65,Fisher67,Kadanoff67,Heller67,Milosevic76,Lipa96}. Response functions associated with the dynamics of manifestly scale-invariant soft excitations in a quantum system, such as X-ray absorption in a metal\cite{Bearden35,Paratt36,Paratt59,Nozieres69} or the spectral function of a Tomonaga-Luttinger liquid (TLL)\cite{Tomonaga50,Luttinger63,Bockrath1999,Auslaender2002,Jompol2009}, likewise display power laws around zero momentum. Recently, the possibility of invariance emerging even in the presence of a length scale was predicted by the newly-developed nonlinear-Luttinger-liquid theory for a one-dimensional (1D) quantum fluid at finite energy and momentum \cite{Imambekov09}. Here, the length scale is determined by the particle's wavelength.

In 1D, effects of electron-electron interactions are amplified strongly, structuring free electrons into collective excitations. These excitations are charge and spin density waves within the TLL theory, which approximates the electron dispersion relation with a linear energy-momentum dependence. The TLL spectral function (which gives the probability of finding an electron with a particular energy and momentum) is zero at energies below the dispersion of the collective mode, and follows a power law in energy measured from the threshold defined by that spectrum; the corresponding exponent depends on the interaction strength\cite{Meden92}.
To go beyond the linear approximation of the energy-momentum dependence and take into account the parabolicity of the dispersion of free electrons, the mobile-impurity model was developed\cite{Imambekov09b}, leading to a non-linear hydrodynamic theory that extends the low-energy universality of a TLL to finite energy, corresponding to excitations from far below the Fermi energy to just above it. Also, the exponent becomes momentum-dependent through a finite curvature of the spectral-edge dispersion that changes with the momentum, defining the momentum dependence as a unique feature of the nonlinear hydrodynamics in 1D.
For electrons (fermions with spin 1/2), this dispersion is close to parabolic\cite{Tsyplyatyev14} and the mobile-impurity model with spin and charge degrees of freedom predicts an essential dependence of the threshold exponent on momentum away from the Fermi points\cite{Schmidt10}. 

Here, we probe the electron dispersion and threshold exponents experimentally by measuring the momentum- and energy-resolved tunnelling between neighbouring 1D and 2D systems, formed within the two layers of a GaAs/AlGaAs double-quantum-well heterostructure. We find the first experimental evidence for the new type of power law in the spectral function. The observed momentum dependence of the power law in the high-energy region matches the theoretical predictions\cite{Schmidt10,Tsyplyatyev14}, supporting not only the 1D theory of interacting particles beyond the linear regime but also the existence of a new type of universality that emerges at finite energy and momentum. This result is a significant stepping stone towards a systematic understanding of a wider variety of many-body systems, ranging from quantum optics to high energy and solid-state physics.

\begin{figure*}[t!]
\includegraphics[width=0.75\textwidth]{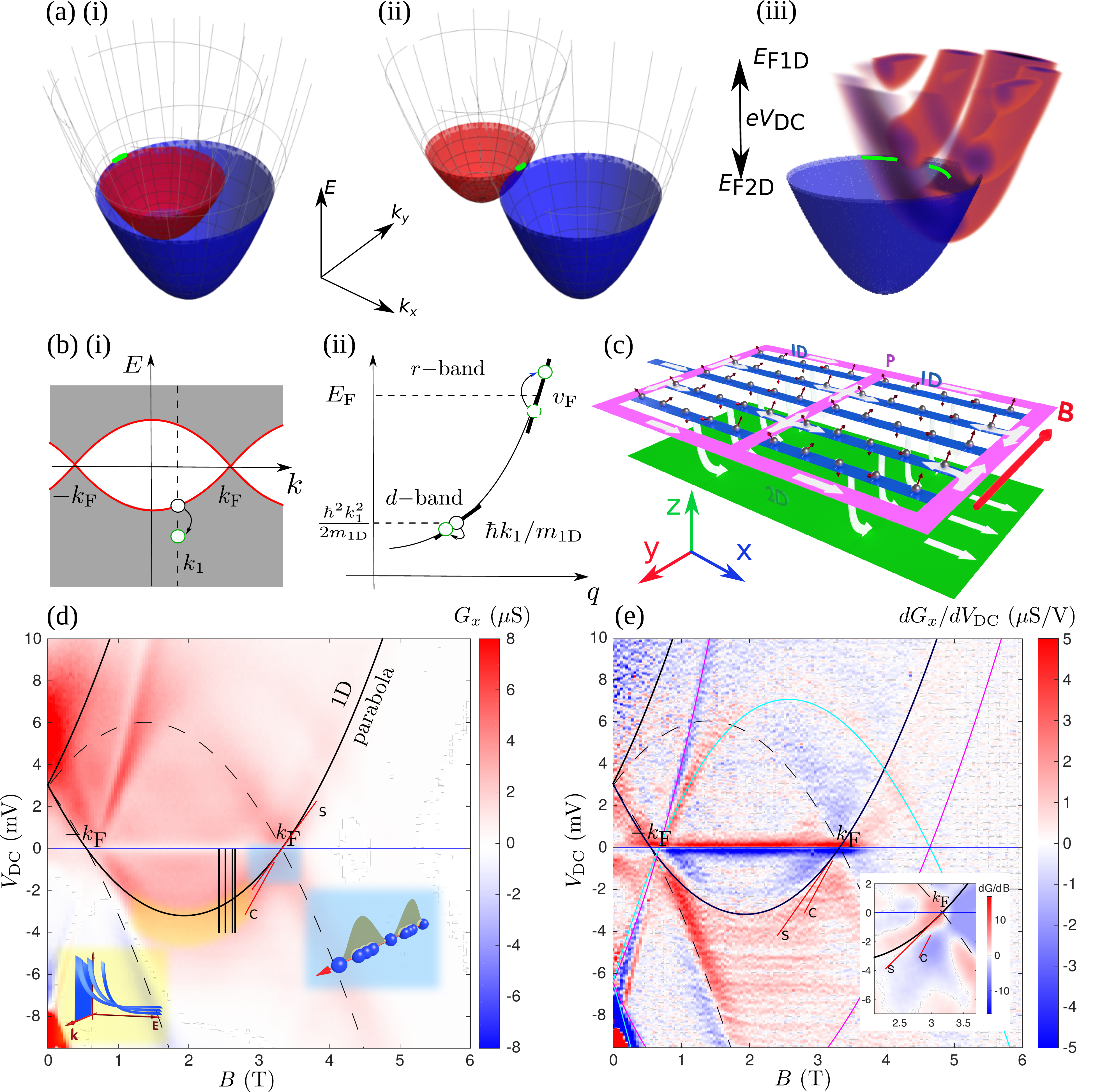}\\
\caption{\label{FigureOverview}\manualsublabel{fig:specfns}{a}{FigureOverview} (i),(ii) Overlap of spectral functions of two 2D layers as a function of momentum $(k_x,k_y)$ and energy $E$---current only flows from occupied states to empty states where the spectral function is not negligible. Magnetic field $B$ displaces one paraboloid to the right. Fermi circles touch on the inside/outside, probing states of the red paraboloid near $-k_{\rm F}$/$k_{\rm F}$ at the Fermi energy $E_{\rm F}$. (iii) 2D system (blue) probing multiple 1D subbands (red), for finite $V_{\rm DC}$. \manualsublabel{fig:replicas}{b}{FigureOverview} (i) Dispersion of an interacting 1D system. White is kinematically forbidden region (see explanation in text), grey is continuum of many-body excitations. Thick red line is border between the two regions. States on the border correspond to removing a single particle (black circle at $k_{1}$) from the many-particle state, and a higher-energy excitation described by mobile-impurity model is marked by a green circle. (ii) Splitting of fermionic dispersion into two subbands, one for heavy hole with velocity $\hbar k_{1}/m_{\rm 1D}$ and one for excitations around $E_{\rm F}$ with velocity $v_{\rm F}$. Green circles are constituent parts of many-body excitations in nonlinear regime (see explanation in text). \manualsublabel{fig:tunnel}{c}{FigureOverview} Schematic of current flow: current enters top layer under mid-gate MG at left, flows into 1D wires (blue) via parasitic regions (magenta, labelled P), and tunnels to lower, 2D, layer and out to a contact at right. \manualsublabel{fig:overview}{d}{FigureOverview} Overview of conductance $G={\rm d}I/{\rm d}V$ \textit{vs} $B$ ($\propto$ momentum) and voltage $V_{\rm DC}$ ($\propto$ energy $eV_{\rm DC}$) (Sample A). Yellow shading shows region along which $G$ is enhanced, inset shows how it might decay with energy, differently for each momentum $k$. Vertical black lines show cuts along which fitting is carried out. Blue shaded region shows where spin-charge separation is visible, inset illustrates charge (dots) and spin (shaded) waves. \manualsublabel{fig:dGdV}{e}{FigureOverview} Differential ${\rm d}G/{\rm d}V_{\rm DC}$ of raw data used for (d) to show a replica above $k_{\rm F}$ and separate spin and charge lines labelled S and C, respectively, and parasitic signal (near magenta and cyan lines). Inset shows ${\rm d}G/{\rm d}B$ around S and C.}
\end{figure*}
\begin{figure*}[t!]
\includegraphics[width=0.9\textwidth]{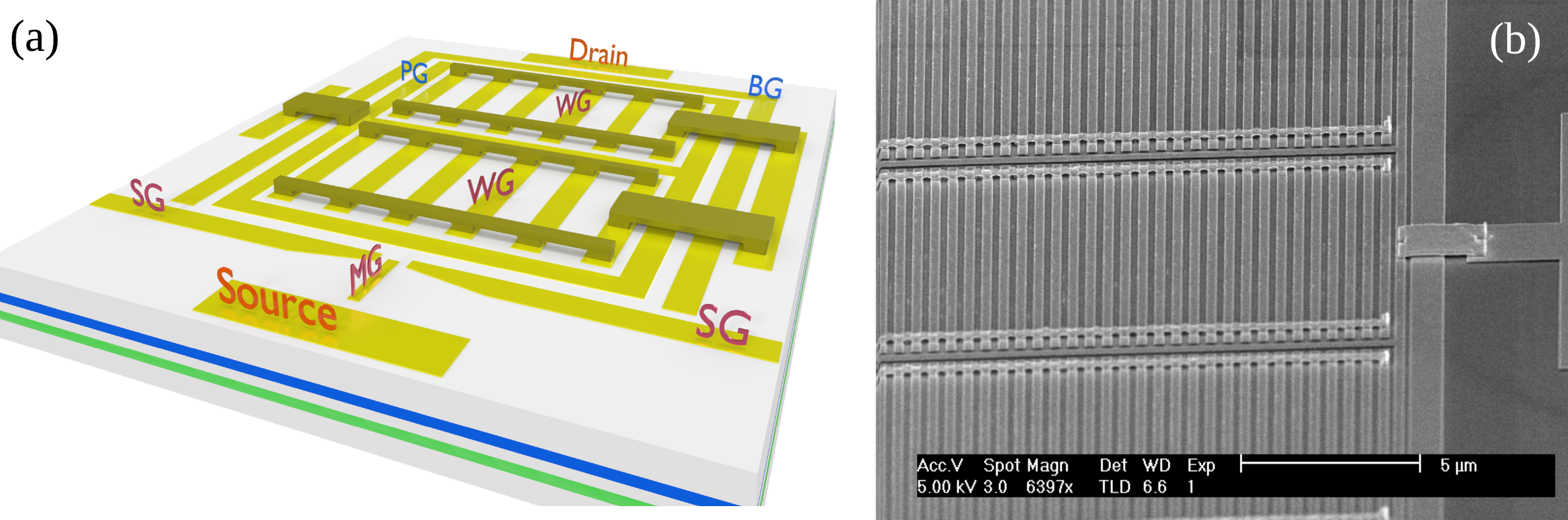}\\
\caption{\label{FigureDevice}\manualsublabel{fig:device3D}{a}{FigureDevice} Illustration of the 1D-2D tunnel device. The control gates are split gate (SG) and mid-gate (MG) (depleting the lower layer but not the top, to inject current into the upper layer), gates defining 1D wires (WG), gate over the parasitic region (PG), and barrier gate (BG) confining the upper layer. Gates are not drawn to scale. In reality they consist of a three-column array of over 500 wires, each of length \SI{18}{\mu m} and gate width \SI{0.1}{\mu m}, with \SI{0.18}{\mu m} separation (device A) and \SI{10}{\mu m} and gate width \SI{0.3}{\mu m}, with \SI{0.2}{\mu m} separation (device B). \manualsublabel{fig:SEM}{b}{FigureDevice} Scanning electron micrograph of a device, showing the air bridges.}
\end{figure*}
\begin{figure*}[t!]
\centering
\includegraphics[width=\textwidth]{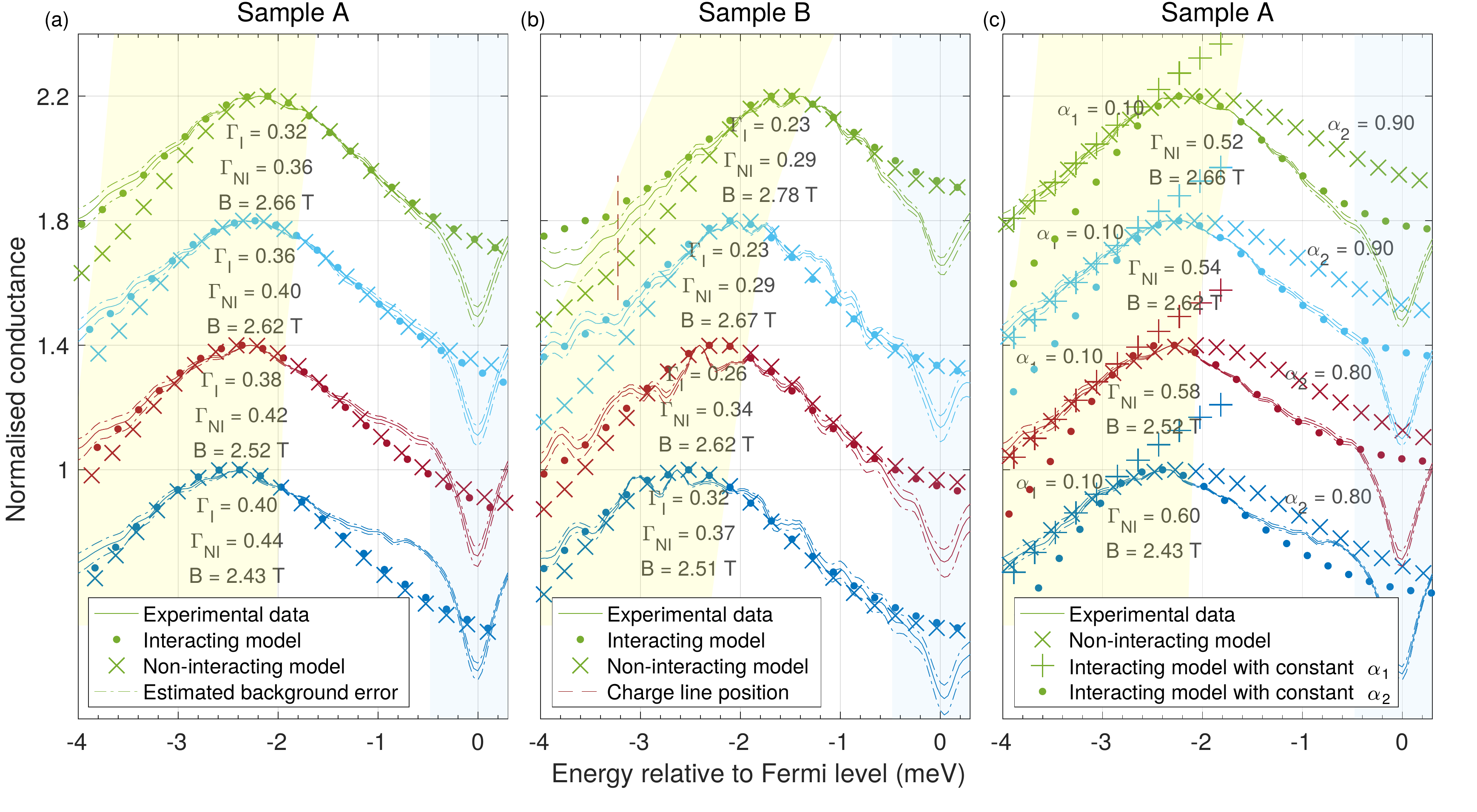}
\caption{\label{FigureG}Fits to the conductance for \manualsublabel{fig:G_devA}{a}{FigureG} sample A and \manualsublabel{fig:G_devB}{b}{FigureG} sample B, normalised to their peak values and shifted vertically for clarity, for model 1 (crosses) and model 2 (dots). The dashed lines show the confidence interval of the measurement data within estimated error of the background conductance subtraction. The broadening fitting parameters for the non-interacting and interacting models ($\Gamma_{\rm NI}$ and $\Gamma_{\rm I}$, respectively) are shown, in units of meV, together the $B$ field at which the data were taken. Sample A (wires \SI{18}{\mu m} long) was measured at \SI{50}{mK} in a He$^3$/He$^4$ dilution refrigerator, while sample B (\SI{10}{\mu m} long) was measured at \SI{330}{mK} in a He$^3$ cryostat.
\manualsublabel{fig:G_otherfits}{c}{FigureG} Fits using other theories, or matching different parts of the data using different constant exponents $\alpha_1$ and $\alpha_2$, as labelled (using data for sample A).}
\end{figure*}
Our devices contain two closely spaced two-dimensional electron gases (2DEGs). The upper 2DEG is depleted into 1D channels by  Schottky gates. A small current flows between the two layers when a suitable magnetic field and DC bias are applied to the device, as illustrated in Fig.~\ref{fig:tunnel} and Fig.\ \ref{FigureDevice}. This current is due to quantum tunnelling and occurs when the unoccupied states of one system align with the occupied states of the other. The alignment of the dispersions is affected by the DC bias $V_{\rm DC}$ and the in-plane magnetic field $B$ perpendicular to the 1D wires, which induce energy and momentum offsets respectively between the two dispersions (see Methods). In Fig.\,\ref{fig:specfns}(i) and (ii), spectral functions of two 2D systems are shown, to illustrate the offsets in $k$ for which significant tunnelling occurs, for $V_{\rm DC}=0$. In Fig.\,\ref{fig:specfns}(iii), the 2D system is shown probing the more complex 1D spectral function (red) with multiple 1D subbands, for finite $V_{\rm DC}$. Fig.\,\ref{fig:overview} shows an overview of such a measurement, where conductance through the sample is measured as a function of energy ($\propto$\,$V_{\rm DC}$) and momentum ($\propto$\,$B$). The conductance peaks form a set of intersecting parabolae, which correspond to the dispersions of each system. The parabolae corresponding to the 1D (2D) system are shown as solid (dashed) lines.


The Luttinger model is only applicable in a small range of excitation energies about the Fermi energy $E_{\rm F}$, which we define as the energy of the highest occupied electron state relative to the bottom of the 1D subband. This corresponds to $V_{\rm DC}=0$. As we can measure excitations at all energies including far above and below $E_{\rm F}$, the predictions of the new nonlinear TLL model can be tested. We have previously observed signs of another nonlinear theory, a hierarchy of modes, which predicts that the spectral power in the many-body continuum varies inversely with the system length\cite{Tsyplyatyev2015}, in the form of `replicas' of the main parabolic dispersion above $E_{\rm F}$: at higher fields above that corresponding to the Fermi momentum $\hbar k_{\rm F}$ \cite{Tsyplyatyev2015,Tsyplyatyev2016} and in the principal region between the $\pm k_{\rm F}$ points (the crossing points labelled $\pm k_{\rm F}$ at $B\sim\SI{0.55}{T}$ and $\sim\SI{3.3}{T}$ in Fig.~\ref{fig:overview}) in short wires \cite{Moreno2016}. Now we turn our attention to the region below the bottom of the 1D parabola, and make detailed fits of the data to models. We find that the tunnelling conductance is significantly enhanced over that predicted by the non-interacting model and cannot be fitted by a model with a simple, momentum-independent power-law dependence, in agreement with the new mobile-impurity model.

Our samples also show the same effects as reported in previous works: (1) Power-law suppression of the tunnelling conductance around zero DC bias (zero-bias anomaly, ZBA), originating from a type of orthogonality catastrophe. This has been seen in carbon nanotubes \cite{Bockrath1999}, and by tunnelling both between two 1D wires \cite{Auslaender2002} and by us between 1D wires and a 2DEG\cite{Jompol2009}. (2) Separation of spinon and holon excitations \cite{Auslaender2002,Jompol2009} close to the Fermi momentum $\hbar k_{\rm F}$). (3) Replica of the principal 1D dispersion just above $k_{\rm F}$, see Fig.~\ref{fig:dGdV}.

\begin{figure}[t]
\centering
\includegraphics[width=0.5\columnwidth]{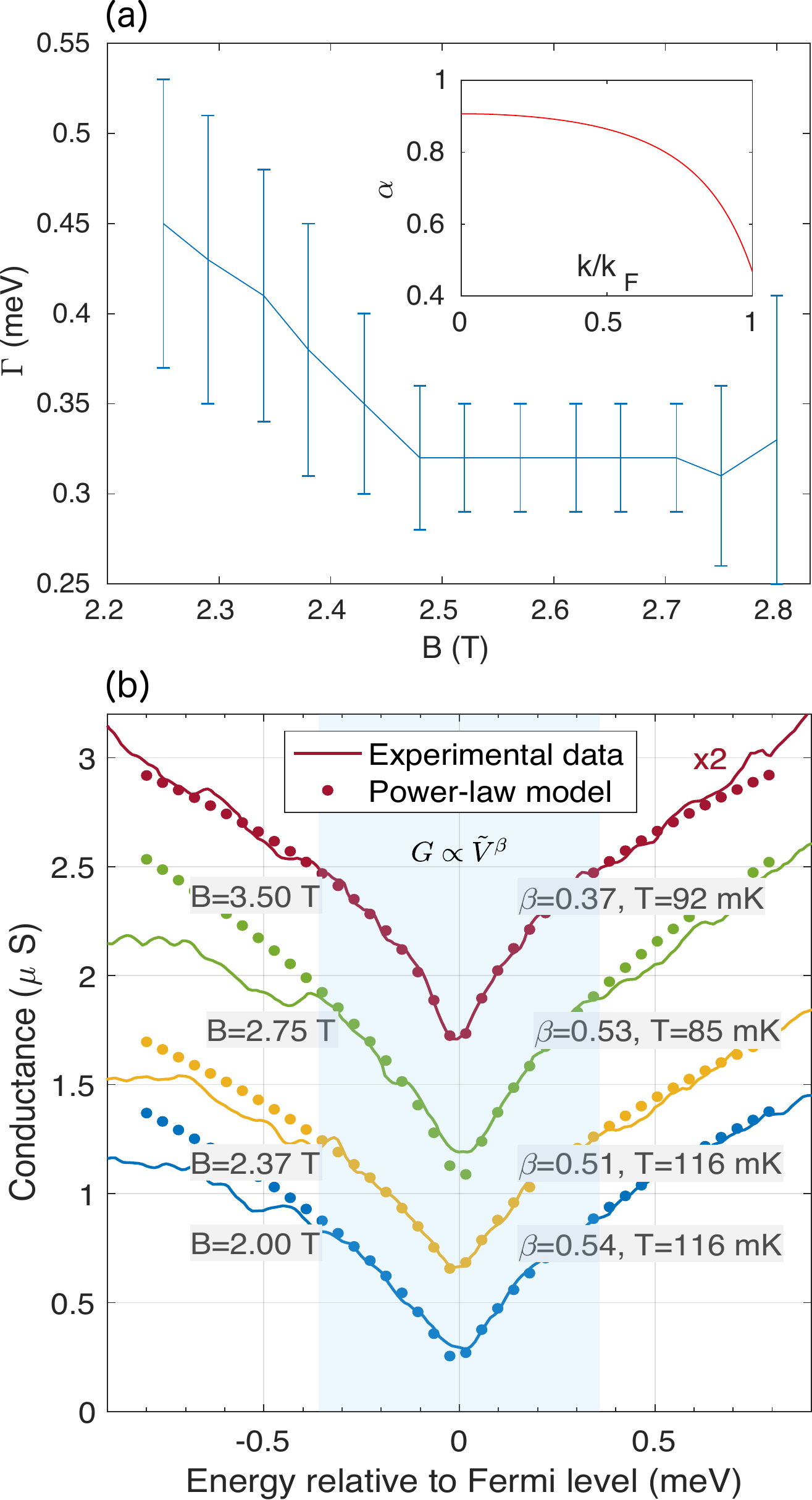}
\caption{\label{FigureGammaAndZBA}\manualsublabel{FigureGamma}{a}{FigureGammaAndZBA} Dependence of the broadening fitting parameter $\Gamma$ on field $B$, for the full interacting calculation (model 2). The inset shows how $\alpha(k)$ varies with $k$, from Eq.~\ref{eq:def_alpha}.  
\manualsublabel{FigureZBA}{b}{FigureGammaAndZBA} Power-law fit of the ZBA for the first sample. The conductance is modelled to be of the form $G = a\tilde{V}^{\beta}$, where $\tilde{V}=\sqrt{V_{\rm AC}^2+V_{\rm DC}^2+(3k_{\rm B}T/e)^2}$. The extracted values of $\beta$ are similar to those obtained originally for carbon nanotubes \cite{Bockrath1999, Kane97} in the end-tunnelling regime and in our previous work \cite{Jompol2009}.
The AC excitation voltage was $V_{\rm AC}=5\,\mu$V. Each sweep has been offset by \SI{0.4}{\mu S} consecutively, for clarity.
}
\end{figure}

The region of interest in this paper (shown shaded yellow in Fig.~\ref{fig:overview}) is below the bottom of the 1D dispersion where a momentum-dependent power law is predicted\cite{Imambekov12}. In order to remove the conductance contribution from a parasitic tunnelling region (magenta and cyan parabolae at high field in Fig.~\ref{fig:dGdV}), a second set of data was measured under identical conditions, but with the 1D wires just past pinch-off, such that the data contain only conductance due to parasitic tunnelling. After subtraction of these data, sharp features of the p-region (the magenta and cyan parabolae) are still noticeable owing to a slight difference in the carrier densities. However, in the region of interest, the conductance varies very slowly with $B$, so the slight variation in density is negligible. The data with the parasitic contribution removed are compared with calculations based on three models that differ in the form of the 1D spectral function: (1) without the power law (or other effect) arising from interactions, (2) with a momentum-dependent power law from interactions, and (3) with a momentum-independent power law.

Model 1 contains a single parameter $\Gamma_{\rm NI}$, the width of the disorder-broadened spectral function. 
The mobile-impurity model for electrons with Coulomb interactions (model 2) predicts the threshold exponents in terms of the curvature of the spinon mode in the nonlinear regime\cite{Schmidt10}. The spectral function of a single 1D subband in the hole sector, $A_1\left(k_x,E\right)\propto 1/\left|E-\varepsilon(k_x)\right|^{\alpha(k_x)}$, is measured directly in our experiment. For a parabolic dispersion\cite{Tsyplyatyev14}, $\varepsilon(k_x)=\hbar^2\left(k_x-k_{\rm F}\right)^2/\left(2m_{\rm 1D} K_s\right)$, the exponent is (see details of the calculation in Supplementary Information, section 2)
\begin{equation}
 \alpha\left(k_x\right)=1-\frac{ \left(1-C\left(k_x\right)\right)^2}{4 K_{\rm c}}-\frac{K_{\rm c} \left(1-D\left(k_x\right)\right)^2}{4}, \label{eq:def_alpha}
\end{equation}
where the momentum-dependent parameters are $C\left(k\right)=\left(k^2-k^2_{\rm F}\right)\allowbreak/\left(k^2/K_{\rm s}-k_{\rm F}^2 K_{\rm s}/K_{\rm c}^2\right)$ and $D\left(k\right)=\left(k-k_{\rm F}\right)\allowbreak\left(k_{\rm F}/K_{\rm c}^2+k/K_{\rm s}^2\right)\allowbreak/\left(k^2/K_{\rm s}^2-k_{\rm F}^2/K_{\rm c}^2\right)$. By renormalisation-group arguments, $K_{\rm s}=1$ in our experiment\cite{Giamarchi} and we use $K_{\rm c}<1$ as a fitting parameter in both the linear and nonlinear regimes.
In addition to these interaction effects, we include also the effect of disorder-induced broadening (of width $\Gamma_{\rm I}$), which smears the threshold singularities to become maxima of the spectral-function energy dependence.
Model 3 is a simpler one for comparison, where the dependence on momentum and $K_{\rm c}$ is replaced with a constant-valued $\alpha$ so that the exponent is momentum-independent.

The models are evaluated as functions of $V_{\rm DC}$ and $B$. Since the calculation is too time-consuming for automated fitting, we calculate the conductance for a range of values of each parameter, and compare the model to the experimental data visually in order to determine the parameters that best fit the experiment (Fig.\,\ref{FigureG}). In this procedure, we ignored the ZBA associated with strong violation of the momentum conservation in tunnelling (see Fig.\ \ref{FigureZBA}, where the ZBA is fitted to the formula used in \citep{Jompol2009}), and concentrated on more negative bias values, corresponding to the range including the conductance maximum and beyond (shaded yellow in Fig.\ \ref{FigureG} and Fig.~\ref{fig:overview}). The value of $K_{\rm c}<1$ affects the skewness of the conduction peak. We find $K_{\rm c}=0.70 \pm 0.03$ best matches the conductance peaks for both samples and all fields. Fig.\,\ref{FigureGamma} shows the range of acceptable $\Gamma_{\rm I}$ values \textit{vs} $B$, for model 2. Between $2.5$ and $2.7$\,T, $\Gamma_{\rm I}$ is at a minimum and roughly constant, as here we avoid the 2D parabola, the localised states at the bottom of the 1D parabola (on the left, see Supplementary Information, section 1, for a discussion of their effects) and the charge line (on the right). The non-zero $\Gamma_{\rm I}$ is largely caused by monolayer fluctuations in the barrier thickness, which give a spread of subband energies of this order. We concentrate on this region because large $\Gamma_{\rm I}$ obscures the effect of $K_{\rm c}$ on the conductance line-shape. We choose representative cuts through the data (shown as black vertical lines in Fig.\,\ref{fig:overview}). The conductance is shown schematically in the lower-left inset, and its differential (without background subtraction) in Fig.\ \ref{fig:dGdV}. The position of the peak corresponds to the 1D parabola. The densities of the two layers are determined from the crossing points (labelled $\pm k_{\rm F}$) and are used to calculate $E_{\rm F}$.

In Fig.~\ref{FigureG}, calculations of the 1D--2D tunnelling conductance using interacting and non-interacting models are compared with our experimental data for two samples (A and B). Cuts through the data are shown as curves for several magnetic fields (marked with black lines in Fig.~\ref{fig:overview}). The data and calculations are normalised by their maximal values. The models ignore any second subband (visible at low fields for sample A as an enhancement close to zero bias). However, they should explain the conductance at and around the maximum, including a relatively wide region at higher negative biases (shaded yellow, Fig.\ \ref{FigureG}) corresponding to the grey continuum of the many-body excitations in Fig.~\ref{fig:replicas}(i). The non-interacting calculation (model 1, crosses) gives a sharp drop to the left of the peak, as there are no states below the parabolic dispersion. However, the full interacting calculation (based on model 2 and exemplified by the top-most fit presented in Fig.\,\ref{FigureG}(a)) predicts exactly this, an enhancement of tunnelling there, as it allows multiple many-body excitations to be created. An example of such an excitation is marked by the green circle in Fig.\ \ref{fig:replicas}(i), composed of a hole deep below $E_{\rm F}$ ($d$-band in Fig.\ \ref{fig:replicas}(ii)) and a number of Luttinger-liquid modes around $E_{\rm F}$ ($r$-band). This predicts a power-law dependence $\alpha$ on energy away from the dispersion relation/band, where $\alpha$ is, remarkably, a function of momentum. Note that, in calculating the tunnelling conductance, one convolves the 1D and 2D spectral functions over all momenta and energies, so the variation of $\alpha$ with $k_x$ must be included and the result at any value of DC bias includes a range of $\alpha$. Model 3 dispenses with this momentum dependence (see Fig.~\ref{fig:G_otherfits}). Neither a fit with the maximum correctly normalised (dots) nor one with the tail aligned to the data ($+$ symbols) matches the data at all well, as they deviate immediately from the data on one or other side of the peak. This shows that the momentum dependence of $\alpha$ is required to get a good fit. This figure also shows another attempt to fit the data using the non-interacting model ($\times$ symbols), where  $\Gamma$ is increased to match the tail, but this clearly gives an unacceptably slow decay to the right of the maximum.

Note that we observe a good fit to the nonlinear theory in both samples studied, with different wire lengths, and at different temperatures. In the Supplementary Information we consider, and exclude, other possible causes of the enhancement of tunnelling conductance below the 1D subband edge.

Another indication of electron-electron interactions is the effective mass $m_{\rm 1D}$ that we observe for the 1D parabola. For the calculated peaks to line up well in energy with the data in Fig.~\ref{fig:G_devA}, we find that $m_{\rm 1D}\sim 0.92m^*$, where $m^*$ is the 2D effective mass. Note that, while there is an uncertainty in the exact tunnelling distance, and hence in the conversion factor between $B$ and momentum, we use the 2D parabolae (measured by the 1D system or the parasitic region) to calibrate the distance. The observed difference in masses is due to non-equal contributions of the interactions in different dimensions to  renormalisation of the free particle mass.


In conclusion, we have studied experimentally the decay of the tunnelling current below the bottom of the 1D subband. The conductance $G$ decays more slowly than predicted by the non-interacting theory, or by interacting theory that includes only a fixed power law $\alpha$. A good fit, however, is obtained using a power law that depends on momentum, as predicted by the recent theory. This appears to be the first example of an interaction-driven variable power law.

\subsection{Methods}

\paragraph{Experimental details} \sloppy The spectrometer device is made with an MBE-grown GaAs/Al$_{0.33}$Ga$_{0.67}$As heterostructure with two parallel quantum wells \SI{100}{nm} beneath the surface, the upper one \SI{18}{nm} wide and the lower one \SI{18}{nm} wide, separated by a \SI{12}{nm} tunnel barrier, giving a \SI{32}{nm} centre-to-centre distance $d$. Fig.~\ref{fig:device3D} is an illustration of the device structure. An array of identical fine-feature wire gates are fabricated onto a Hall bar by electron-beam lithography. They are inter-connected by air bridges, which are used to supply a negative voltage to deplete the upper 2DEG layer into 1D channels while leaving the lower 2DEG undisturbed. A split gate/mid-gate arrangement is positioned near the source contact of the Hall bar while a barrier gate is placed on the opposite side near the drain contact. The split gates are used to pinch off both the upper and lower 2DEG layers, while the mid gate supplies a positive voltage to induce a channel in the upper 2DEG from where electrons can flow into the array. The split gate/mid-gate combination ensures electrons from the source Ohmic contact may only enter the 1D array via the upper layer. On the opposite side, the barrier gate just pinches off the upper 2DEG. Electrons may only leave the lower 2DEG via the drain Ohmic contact. As illustrated by Fig.~\ref{fig:tunnel}, electrons must tunnel between the 2DEG layers in order to travel between the source-drain contacts.

Fig.~\ref{fig:tunnel} shows that there are two of regions where electron tunnelling occurs in the upper 2DEG: (1) The 1D channels defined by the wire gate array (marked blue in figure). (2) The regions surrounding the wire gate array, which allows electron flow into the array (marked purple in figure). We detect tunnelling from both regions in the experiment. Analysis is focused on the tunnelling from the array. The strength of confinement of the 1D channels can be controlled by the wire gates. The array, which contains 500 repeating units, provides a large total tunnelling area to provide a strong conductance to be measured. Tunnelling from the second region is parasitic---it cannot be eliminated due to device design. A control gate was fabricated on top of the parasitic region in order to provide control over the electron density of the region.


The experiment was carried out at $T<\SI{100}{mK}$ in a $^3$He/$^4$He dilution refrigerator (sample A) or at $T\sim\SI{330}{mK}$ in a $^3$He cryostat (sample B). Device conductance was measured in a two-terminal phase-sensitive setup, where a small AC voltage was applied as the source-drain bias and the current response measured by a lockin amplifier. 
The wire-gate voltage is chosen to be negative enough that only the 1D states in lowest sub-band are populated, though for smaller voltages up to three 1D subbands can be observed clearly. The conductance across the sample was measured as the DC bias was swept and the magnetic field incremented.

The tunnelling current between the two 2DEG layers is given by \cite{Altland1999}:
\begin{multline}
  I \propto \int\mathrm{d}\mathbf{k} \mathrm{d}E \big[f_T\left(E-E_{\rm F1D}-eV_{\rm DC}\right) - f_T\left(E-E_{\rm F2D}\right)\big]\\
  \times A_1(\mathbf{k},E) A_2(\mathbf{k}+ed\left(\mathbf{n}\times\mathbf{B}\right)/\hbar,E-eV_{\rm DC}), \label{TunnelCurrent}
\end{multline}
where 
$e$ is the electron charge,
$f_T(E)$ the Fermi-Dirac distribution function, $d$ is the spatial separation between the two layers of 2DEGs, $\mathbf{n}$ is the unit normal to the surface, $\mathbf{B}=-B\hat{\mathbf{y}}$ is the magnetic-field vector (magnitude $B$), $\hat{\mathbf{y}}$ is the unit vector in the $y$-direction, $A_1$ and $A_2$ are the spectral functions of the 1D and 2D systems, respectively, and their corresponding Fermi energies are $E_{\rm F1D}$ and $E_{\rm F2D}$. According to Eq.~(\ref{TunnelCurrent}), the tunnelling current between the two layers is proportional to the overlap integral of the spectral functions of the two layers. 
We can induce an offset $eV_{\rm DC}$ in the Fermi energies between the two layers by applying a DC bias $V_{\rm DC}$. A momentum offset can be induced by a magnetic field of strength $B$ parallel to the 2DEG layers (as shown in Fig.\,\ref{fig:tunnel}). Assuming that the field direction is along the $y$-axis, the vector potential is equal to $\mathbf{A} = (zB,0,0)$ in the Landau gauge, and the Lorentz force shifts the momentum of the tunnelling electrons in the $x$-direction by ${\mathbf{p}=\hbar\mathbf{k}=(-edB,0,0)}$. At low temperatures, the Fermi-Dirac distributions can be approximated by a Heaviside step function $\theta(x)$. 

\paragraph{Modelling details} The tunnelling rate is proportional to the spectral function, which gives the probability density to find an electronic state at a given point of energy-momentum space. The spectral function can be obtained via a Fourier transform of the real-space Green function \cite{AGD}. The latter is expressed in terms of electron wave functions.
In a free 2D space, the electron wave function is a plane wave, so the spectral function is a delta function: ${A_2(\mathbf{k},E) = \delta \left(E - \varepsilon(\mathbf{k})\right)}$, where $\varepsilon(\mathbf{k})$ is the dispersion relation $\hbar^2(k_x^2+k_y^2)/2m$. To account for disorder broadening, the spectral function is convolved with a Lorentzian function with spread $\Gamma$: 
\begin{equation}
  A_2(\mathbf{k}, E, \Gamma) =\frac{\Gamma}{\pi} \frac{1}{ \Gamma^2 + \left(E - \frac{\hbar^2(k_x^2+k_y^2)}{2m^*} \right)^2}.
  \label{ModelCurrent}
\end{equation}

Experimentally, the gate-induced 1D channels have finite transverse confinement potentials (instead of being infinitely narrow and hence having infinite subband spacing). For this reason, the 1D spectral function depends on the transverse dimension, $k_y$. The confinement potential can be treated as a parabolic quantum well, whose electron wave function is given by a quantum-harmonic oscillator solution \cite{Kardynal97}. The confinement results in energy levels known as 1D subbands. The spectral function of the 1D system is given by the Fourier transform of the wave functions summed over all subbands which contribute to conduction (i.e.\ below $E_{\rm F}$), and convolved with a Lorentzian function to account for broadening. Without considering the effects of interactions, the 1D spectral function is:
\begin{equation}
  A_{1\mathrm{non-int}}(\mathbf{k}, E, \Gamma) = \sum_{n}\frac{ \Gamma}{\pi}\frac{H_n(k_y a)e^{-(k_y a)^2}}{  \Gamma^2 + \left(E -E_n\left(k_x\right)\right)^2 },
\end{equation}
where $n$ is the 1D subband index, $a=m_{\rm 1D}\omega/\hbar$ is a finite width of the wire in the y-direction, $E_n\left(k_x\right)= k_x^2/(2m_{\rm 1D})+\hbar\omega\left(n+1/2\right)$ is the parabolic dispersion of each subband, $\hbar\omega$ is the energy spacing of the subbands, and $H_n(x)$ are the Hermite polynomials. As was shown in the main text, the 1D spectral function derived from the mobile-impurity model is $A_1(k_x,E)\propto 1/|E-\varepsilon(k_x)|^{\alpha(k_x)}$ (see details in the supplementary material) and the momentum dependence of the exponent is given by Eq.~(\ref{eq:def_alpha}). The 1D spectral function that includes the effects of interactions is therefore:
\begin{multline}
	A_{1\mathrm{int}}(\mathbf{k},E,\Gamma) = \displaystyle\int\limits_{-\infty}^{\infty}\mathrm{d}z\sum_{n} \frac{\theta(E - E_n(k_x)-z)}{(E - E_n(k_x)-z)^{\alpha\left(k_x\right)}}\cdot\\ H_n(k_y a)e^{-(k_y a)^2}\frac{\Gamma}{\pi}\frac{1}{\Gamma^2+z^2},
\end{multline}
where a finite number of 1D subbands $n$ is taken into account.

The integral of (\ref{ModelCurrent}) was evaluated numerically using Mathematica, which gave the tunnelling current across the sample and was used to calculate the conductance after taking the derivative with respect to $eV$. The calculation and the experimental results were normalised to their own conduction peak values and compared in Fig.~\ref{FigureG}.



\bibliography{citations}{}


\begin{itemize}
\item[] {\bf Acknowledgements} This work was supported by the UK EPSRC [Grant Nos. EP/J01690X/1 and EP/J016888/1]. OT\ acknowledges support from the German DFG through the SFB/TRR 49 programme. LG acknowledges support from NSF DMR Grant No. 1603243.

\item[] {\bf Author contributions} Project planning: CJBF, OT, LIG and AJS; MBE growth: IF and DAR; e-beam lithography: JPG; sample fabrication: YJ and MM; transport measurements: YJ, MM, AA, WKT and CJBF; analysis of results and theoretical interpretation: YJ, CJBF and OT.

\item[] {\bf Data availability} Data associated with this work are available at the University of Cambridge data repository (http://dx.doi.org/10.17863/...;).
\item[] {\bf Competing Interests} The authors declare that they have no competing financial interests.
\item[] {\bf Correspondence} Correspondence and requests for materials
should be addressed to C.\ J.\ B.\ Ford (email: cjbf@cam.ac.uk) or O.\ Tsyplyatyev (email: o.tsyplyatyev@gmail.com).
\end{itemize}

\onecolumngrid

\newpage
\section*{Extended Data}
\begin{figure}[h]
\includegraphics[width=0.8\textwidth]{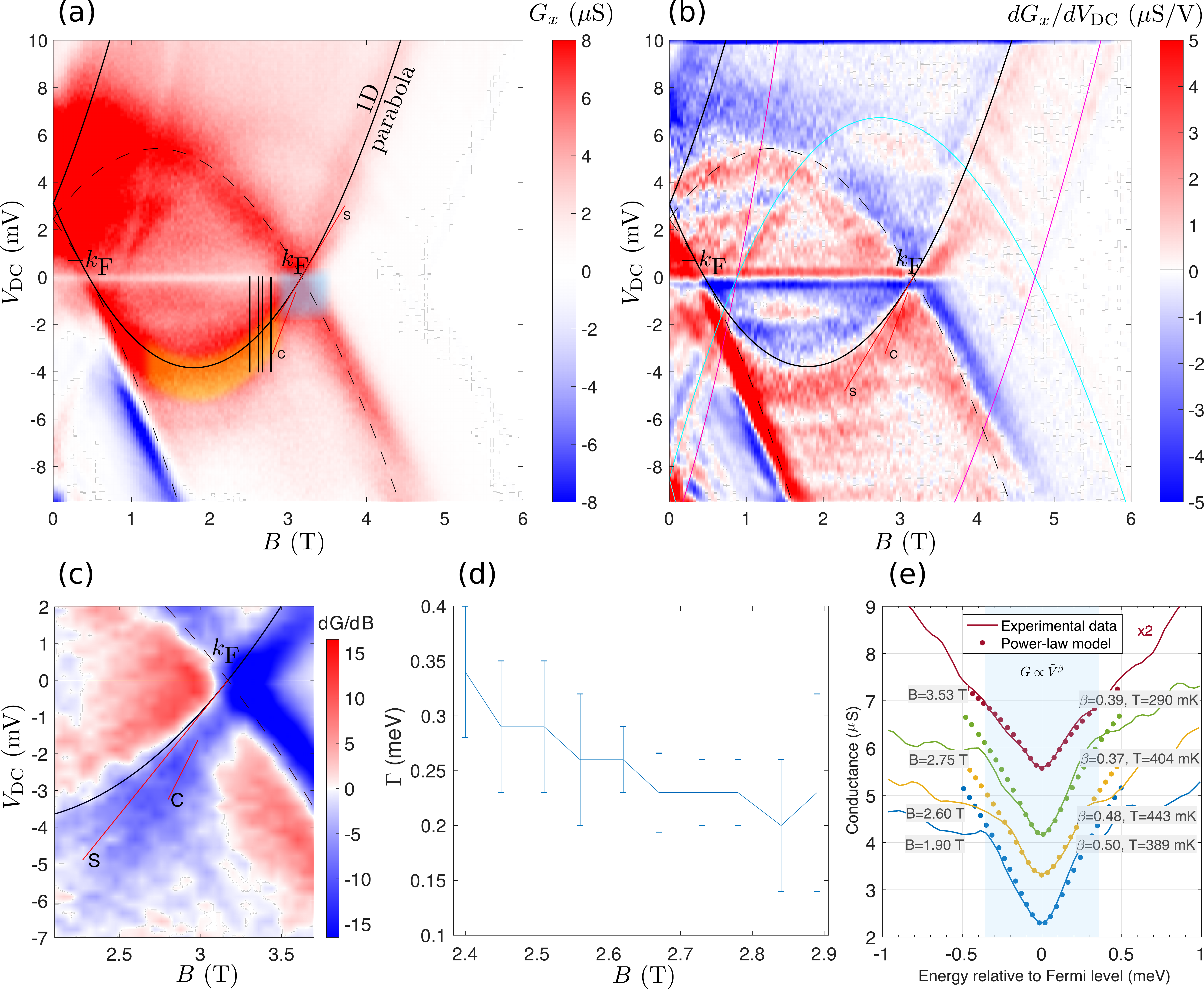}
\caption{Detailed plots of data for sample B, matching those shown in the paper for sample A. (a) Density plot of $G$ (with correction for the parasitic conductance). (b) ${\rm d}G/{\rm d}V_{\rm DC}$ without any correction. (c) Zoom-in on the spin-charge separation region of ${\rm d}G/{\rm d}B$. (d) $\Gamma_{\rm I}$ \textit{vs} $B$. (e) Power-law fits for the ZBA. The sample temperature was about 330\,mK.\label{FigureOverview_2nd_sample}}
\end{figure}

\begin{figure}[!t]
\begin{center}\includegraphics[width=0.8\columnwidth]{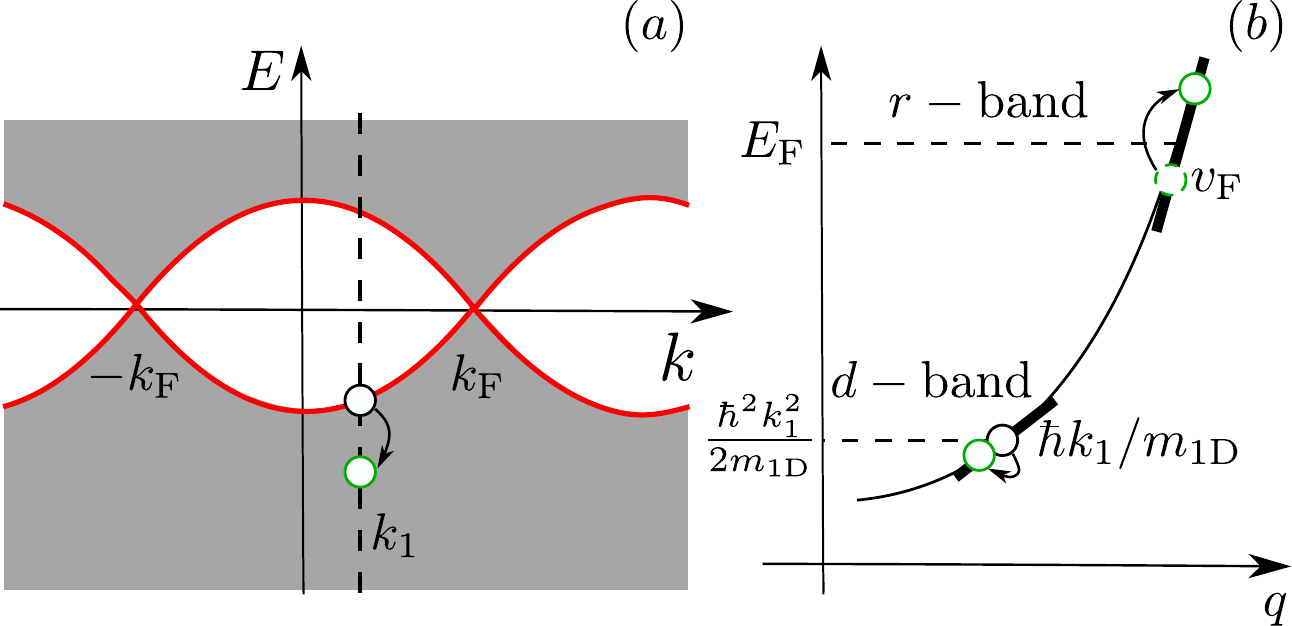}\end{center}
\caption{\label{fig:nL_1D}(a) Dispersion of an interacting 1D system. White is the kinematically forbidden region (see explanation in the text) and grey is the continuum of many-body excitations. The thick red line separates the border between the two regions, the states on the border that form it correspond to removing a single particle (marked by a black circle at $k_{1}$) from the many-particle state, and an excitation described by the mobile-impurity model marked by a green circle. (b) Splitting of the fermionic dispersion into two subbands, one for the heavy hole with velocity $\hbar k_{1}/m$ and one for the excitations around $E_{\rm F}$ with velocity $v_{\rm F}$. The green circles are the constituent parts of the many-body excitations in the nonlinear regime (see explanation in the text).}
\end{figure}
\clearpage

\section*{Supplementary Information}

\subsection{1.~Other possible causes of conductance enhancement}

It is important to exclude other possible causes of the enhancement of tunnelling conductance below the 1D subband edge. We have already removed a background parasitic conductance. In Fig.~\ref{FigureG}, the dashed lines show the conductance where this background has been increased by 20\% (allowing for a slight reduction in area of the parasitic region with gate voltage) or decreased by 20\% (allowing for a remnant of the 1D signal in the background). This variation reflects the uncertainty in the background-elimination process, arising from the slight change in density and area of the p region when the wire gates are pinched off. The enhancement above the conductance predicted by models 1 and 3 is still significant, whereas there is still a good fit to model 2.
As the 1D wires are squeezed, minute lithographic imperfections and occasional impurities will cause fluctuations in the wire width and potential depth. When the wires are nearly pinched off, tunnel barriers may form across a wire, producing localised quantum-dot states. Given that the tunnelling probability across the 12\,nm AlGaAs barrier between the array of 1D wires and the 2D layer is very low, the wires are likely to remain equipotentials despite possibly being broken up into segments in some places. 

The effect of this small disorder on electrons is strongly amplified in 1D according to the one-parameter scaling theory\cite{Abrahams79}, resulting in a fully (Anderson) localised \cite{Anderson58} single-particle spectrum in the thermodynamic limit. For a system of finite length, however, above some energy (mobility) threshold the energy-dependent localisation length exceeds the sample length 
\cite{Altshuler89,Shlyapnikov07}, e.g.\ as was observed directly in cold-atom experiments \cite{Aspect08}. 
In this experiment (see Fig.~\ref{FigureGamma}), $\Gamma_{\rm I}$ increases sharply for $B<2.5$\,T, corresponding to band energies below the mobility threshold ($\simeq 0.18 E_{\rm F}$ above the bottom of the band here),
but above this field we have access to the nonlinear TLL physics in the rest of the 1D band. Note that in the interacting model, the states below the 1D parabola are not single-particle states but many-body states (Fig.~\ref{fig:replicas}(ii)). The disorder below the mobility threshold therefore does not affect the many-body states directly since they are formed by higher-momentum constituents. 

Note that there are some horizontal streaks in Fig.\ \ref{fig:overview} and \ref{fig:dGdV}, which contribute only about 5\% of the overall conductance in most measurements and can be seen as small fluctuations on the curves in Fig.~\ref{FigureG}. They must be caused by tunnelling via localised states where $k_x$ is no longer a good quantum number. As they are seen in a wide range of energies uncorrelated with the 1D band, they may arise from occasional defects or large clusters of donors where the depth of the potential may be as low as 10\,meV below the Fermi level. If such states were the cause of the observed excess in signal which we attribute to interactions (Fig.~\ref{FigureG}), the fact that this is strongly $B$-dependent would require a sharp cutoff in the size distribution of these localised states or dots. The excess changes rapidly between 2.4 and 2.6\,T, and if we estimate this to be when the length of the dot becomes comparable to the magnetic length $l_{\rm B}=\sqrt{\hbar/eB}$, with $l_{\rm B}=16.5$\,nm at 2.4\,T and 15.9\,nm at 2.6\,T, there must be dots of sizes down to 16.0\,nm but far fewer just below this. Hence there is a particular size distribution, which is very unlikely. The small contribution from the observed horizontal streaks is just superimposed on top of the 1D signal and does not affect the model.

\subsection{2.~Modelling the 1D spectral function in nonlinear regime}

In 1D, the theoretical description of the many-body excitations away from the Fermi points is given by the nonlinear TTL theory \cite{Imambekov09}, which we call model 2. It is instructive to consider the original model of interacting fermions first. Any two-body interaction changes the delta-function excitation spectrum of the free particles (centred at the single-particle parabola) into a continuum, since removal of a single particle from the system affects all other particles, by involving their many degrees of freedom. In two and three dimensions this continuum covers the whole energy-momentum space since it is always possible to create an excitation at an infinitesimally small energy at all finite momenta by connecting two points on a circle or on a sphere by a finite vector of length $\left|\mathbf{k}\right|<k_{F}$. The fact that there are only two Fermi points in 1D makes it very special. There is a minimal energy for removing a single particle, with a finite $k$. The process corresponds to taking out just this one particle without touching the rest (shown as a black circle in Fig. \ref{fig:nL_1D}(a)). This leads to a forbidden region on the energy-momentum plane (see white regions in Fig. \ref{fig:nL_1D}(a)), which are separated from the many-body continuum of the excitations by a line that is given by the dispersion of single hole with the minimal energy (see thick red line in Fig. \ref{fig:nL_1D}(a)).

{\it Nonlinear hydrodynamics.} A hole state deep under the Fermi surface is reminiscent of another problem---X-ray scattering in metals, where the deep hole is created by absorption of a high-energy X-ray photon. This system is known to have power-law singularities close to the Fermi level that originates from the interactions between the deep hole and the quasiparticles around $E_{\rm F}$. Although the standard perturbation theory in the interaction for the X-ray problem is divergent, a way of handling these divergences was proposed by Nozi\`{e}res and De Dominicis \cite{Nozieres69} in the form of the heavy-impurity model. This model, consisting of the Fermi-liquid quasiparticles interacting with a localised state deep under the Fermi level, can be diagonalised exactly, accounting for all divergences in all orders of perturbation theory. It predicts power-law behaviour around $E_{\rm F}$. An analogous construction of a mobile-impurity model can be done in 1D starting from the Tomonaga-Luttinger model \cite{Giamarchi},
\begin{equation}
H_{\rm TLL}=\frac{\hbar v_{c}}{2\pi}\int dx\left[K_{c}\left(\nabla\theta_{c}\right)^{2}+\frac{1}{K_{c}}\left(\nabla\varphi_{c}\right)^{2}\right]+\frac{\hbar  v_{s}}{2\pi}\int dx\left[K_{s}\left(\nabla\theta_{s}\right)^{2}+\frac{1}{K_{s}}\left(\nabla\varphi_{s}\right)^{2}\right],\label{eq:HnTTL}
\end{equation}
where $\theta_\alpha$ and $\varphi_\alpha$ are canonically conjugated variables $[\varphi_\alpha(x),\nabla \theta_\beta(x')]=i\pi\delta_{\alpha\beta}\delta(x-x')$ that describe the charge-density wave (CDW) $\alpha=c$ and the spin-density wave (SDW) $\alpha=s$, and $v_\alpha$ and $K_\alpha$ are the Luttinger parameters of these modes that are input parameters of the model. Their values have to be specified for a particular system. Below we consider a strictly one-dimensional system omitting the spatial direction index that we use in the main text, $k=k_x$, for brevity and keep $K_\alpha, v_\alpha$ arbitrary for generality.

A well-defined deep-hole state can be added to $H_{\rm TTL}$ assuming that the hole state is sufficiently far in energy from the Fermi level (see the construction in Fig.~\ref{fig:nL_1D}(a) and (b)) that it is not dynamically created or annihilated by the low-energy excitations. The dispersion of this hole is now an arbitrary input parameter of the model, $\varepsilon\left(k\right)$, describing phenomenologically the dispersion of the spectral edge, see the red line in Fig.~\ref{fig:nL_1D}(a). Its coupling to the CDW of the linear Luttinger liquid is of the density-density type since exchange processes are forbidden by a larger energy difference between the hole band and $E_{\rm F}$, see Fig.~\ref{fig:nL_1D}(b). The mobile impurity does not couple to the spin modes in an unpolarised TLL owing to symmetry of the two spin orientations. The linear Tomonaga-Luttinger model, a mobile impurity, and coupling between them together form the nonlinear TLL model (also called the mobile-impurity model) \cite{Pustilnik06,Khodas07},
\begin{equation}
H_{\rm nTLL}=H_{\rm TTL}+\int dx\,d^{\dagger}\left(x\right)\left(\varepsilon\left(k\right)-i\frac{\partial\varepsilon\left(k\right)}{\partial k}\nabla\right)d\left(x\right)
+\int dx\left(V_{\theta}\nabla\theta_{c}+V_{\varphi}\nabla\varphi_{c}\right)d^\dagger\left(x\right) d\left(x\right),\label{eq:HnLL_spinful}
\end{equation}
where  the fermionic operator $d\left(x\right)$, satisfying $\left\{d\left(x\right),d^\dagger\left(x'\right)\right\}=\delta\left(x-x'\right)$, models a deep hole state, and the coupling constants $V_{\theta}$ and $V_{\varphi}$ are also not independent parameters. They can be related to the linear Luttinger parameter and the dispersion of the spectral edge by considering the velocity of the whole system and the variation of the total energy with the density, which fixes the couplings as \cite{Schmidt10}
\begin{eqnarray}
V_{\theta} & = & \frac{\hbar^2 k}{\sqrt{2}m_{\rm 1D}}-\frac{\partial\varepsilon\left(k\right)}{\sqrt{2}\partial k},\label{eq:Vtheta_spinful}\\
V_{\varphi} & = & \frac{\hbar\partial\varepsilon\left(k\right)}{\sqrt{2}\partial\rho}+\frac{\hbar v_{s}}{\sqrt{2}K_{s}}.\label{eq:Vphi_spinful}
\end{eqnarray}

The validity of the nonlinear TTL model is restricted to the proximity of the spectral edges \cite{Imambekov12}. Moving a deep hole down in energy (for example in the hole sector) away from the the spectral threshold (see the green circle in Fig.~\ref{fig:nL_1D}(a)) requires creation of CDWs with higher and higher energy (see green circles in the r-band in Fig.~\ref{fig:nL_1D}(b)) according to the model in Eq.~(\ref{eq:HnLL_spinful}). However, for sufficiently large energies, the original linear approximation of the Tomonaga-Luttinger model becomes invalid, which also voids the validity of the nonlinear model. Due this constraint the dispersion of mobile impurity in the second term in Eq.~(\ref{eq:HnLL_spinful}) is also linearised around a given momentum $k$ defining a linear impurity subband (see d-band in Fig.~\ref{fig:nL_1D}(b)) and simplifying diagonalisation of the nonlinear model. The nonlinear excitations around the spectral edge, described by the mobile-impurity model in Eq.~(\ref{eq:HnLL_spinful}), are composite many-body states consisting of a deep hole and a relatively small number of CDWs around the around $E_{\rm F}$ (see all green circles in Fig.~\ref{fig:nL_1D}(b)).

{\it Power-law singularities around spectral edges.} The mobile-impurity model in Eq.~(\ref{eq:HnLL_spinful}) can be diagonalised using a unitary rotation in the two-by-two space of the Tomonaga-Luttinger model and the mobile impurity. Then, the expectation values for the observables can be evaluated using the Gaussian integrals over the free fields, as for the linear Tomonaga-Luttinger liquid. The diagonalisation is performed via the $e^{-iU}H_{\rm nTLL}e^{iU}$ rotation, where the rotation matrix can be found in the perturbation-theory analysis from the condition $\left[H,U\right]=0$ as 
\begin{equation}
U=\int dx\left[C_{+}\left(\sqrt{K_{\rm c}}\theta_{\rm c}+\varphi_{\rm c}/\sqrt{K_{\rm c}}\right)+C_{-}\left(\sqrt{K_{\rm c}}\theta_{\rm c}-\varphi_{\rm c}/\sqrt{K_{\rm c}}\right)\right]d^\dagger\left(x\right) d\left(x\right),
\end{equation}
where the coefficients are 
\begin{equation}
C_{\pm}=\frac{\frac{\hbar\left(k-k_{\rm F}\right)}{m_{\rm 1D}\sqrt{K_{c}}}\pm\sqrt{K_{\rm \rm c}}\left(\frac{2}{\pi}\frac{\partial\varepsilon\left(k\right)}{\partial\rho}+\frac{\partial\varepsilon\left(k\right)}{\hbar\partial k}\right)}{2\sqrt{2}\left(\frac{\partial\varepsilon\left(k\right)}{\hbar\partial k}\mp\frac{\hbar k_{\rm F}}{m_{\rm 1D}K_{\rm c}}\right)}, \label{Cpm}
\end{equation}
and the coupling constant from Eqs. (\ref{eq:Vtheta_spinful},\ref{eq:Vphi_spinful}) have already been substituted.

The averages with respect to the free model after the rotation can be evaluated as  integrals over the free bosonic and fermionic fields. We will consider only the spectral function here. It can be defined using the Green function as \cite{AGD}
\begin{equation}
A_1\left(k,E\right)=-\frac{1}{\pi}\textrm{Im}G_{\alpha\alpha}\left(k,E\right)\textrm{sign}E,
\end{equation}
where the real-frequency Green function is a Fourier transform of the time-ordered two-point correlation function, $G_{\alpha\beta}\left(k,E\right)=-i\int dxdt\exp\left(iEt/\hbar-ikx\right)\left\langle T\psi_\alpha\left(x,t\right)\psi_\beta\left(0,0\right)\right\rangle $. Since the model $H_{\rm nTLL}$ is $SU(2)$ symmetric in the absence of a magnetic field the spectral function is the same for both spin orientations and the spin index $\alpha$ will be omitted. The fermionic excitations give the dominant contribution to the spectral function close to the spectral threshold. Thus, its most singular part can be expressed through the correlation function of the mobile impurity operator as \cite{Imambekov09b}
\begin{equation}
A_1\left(k,E\right)\propto\int dtdxe^{i\left(Et/\hbar-kx\right)}\left\langle d^{\dagger}\left(x,t\right)d\left(0,0\right)\right\rangle
\label{eq:A1_theory}
\end{equation}
where the time evolution of the mobile-impurity hole operator is given by the model in Eq.~(\ref{eq:HnLL_spinful}), $d\left(x,t\right)=\exp\left(-iH_{\rm nTLL}t/\hbar\right)d\left(x\right)\exp\left(iH_{\rm nTLL}t/\hbar\right)$, and the expectation value has to be taken also with respect to the whole model in Eq.~(\ref{eq:HnLL_spinful}). 

Integration over the bosonic field in Eq.~(\ref{eq:A1_theory}), in the same way as for the linear Tomonaga-Luttinger model, gives a power-law function in energy,
\begin{equation}
A_1\left(k,E\right)\propto\frac{1}{\left|E\pm\varepsilon\left(k\right)\right|^{\alpha_{\pm}}},
\end{equation}
where the exponent depends on momentum \cite{Schmidt10,Schmidt10b}, 
\begin{equation}
\alpha_{\pm}(k)=\frac{1\mp1}{2}-\frac{1}{2}\left(\frac{\left(2l+1\right)\sqrt{K_{\rm c}}}{\sqrt{2}}-C_{+}-C_{-}\right)^{2}-\frac{1}{2}\left(\frac{1}{\sqrt{2K_{\rm c}}}-C_{+}-C_{-}\right).\label{eq:alpha_spinful}
\end{equation} 
Here $\pm$ refer to the particle (hole) sector on the energy-momentum plane and $l$ is the integer number of translations of the principal region in the momentum variable from $-k_{\rm F}$ to $k_{\rm F}$ by $2k_{\rm F}$. Substitution of the parabolic dispersion for a repulsive spinful fermionic system obtained using Bethe ansatz method \cite{Tsyplyatyev14},
\begin{equation}
\varepsilon(k)=\mu+\frac{\hbar^2\left(k^2-k_{\rm F}^2\right)}{2m_{\rm 1D}K_s},  
\end{equation}
where $\mu$ is the chemical potential,  into Eqs. (\ref{Cpm}) and (\ref{eq:alpha_spinful}) for $l=0$ in the hole sector gives the explicit momentum dependence of the edge exponent quoted in Eq.~(1) of the main text.

\end{document}